\documentstyle[aps,prl,epsf,floats,twocolumn]{revtex}

\begin{document}
\draft

\title{Comment on ``Quantum Phase Transition of the Randomly Diluted \\
Heisenberg Antiferromagnet on a Square Lattice''}

\author{Anders W. Sandvik\cite{present}} 

\address{Department of Physics, University of Illinois at Urbana-Champaign, 
1110 West Green Street, Urbana, Illinois 61802}

\date{\today}
\vskip-2mm

\maketitle{}

%\pacs{PACS numbers: 75.10.Jm, 75.10.Nr, 75.40.Cx, 75.40.Mg}

In \cite{kato}, Kato {\it et al.} presented quantum Monte Carlo results 
indicating that the critical concentration of random non-magnetic sites in the 
two-dimensional antiferromagnetic Heisenberg model equals the classical 
percolation density; $p_{\rm c}=p^* \approx 0.407254$. The data also 
suggested a surprising dependence of the critical exponents on the spin $S$ 
of the magnetic sites, with a gradual approach to the classical percolation 
exponents as $S \to \infty$. This behavior is contrary to well established
notions of universality, according to which the exponents should depend only 
on symmetries and dimensionality. I here argue that the exponents in fact 
are $S$-independent and equal to those of classical percolation. The apparent
$S$-dependent behavior found in \cite{kato} is due to temperature effects 
in the simulations as well as a quantum effect that masks the true asymptotic 
scaling behavior for small lattices.

\begin{figure}
\centering
\epsfxsize=8.4cm
\leavevmode
\epsffile{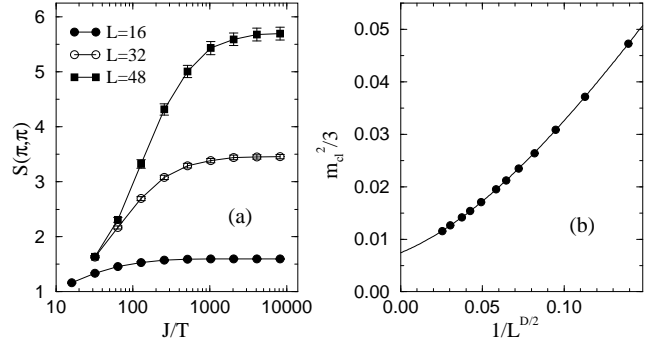}
\vskip1mm
\caption{(a) Temperature dependence of the structure factor for 
different system sizes. (b) Size-dependence of the cluster magnetization
for $L$ in the range 8,...,48.}
\label{fig}
\end{figure}

I discuss simulation results for the $S=1/2$ case obtained using the 
stochastic series expansion method \cite{sse} on $L \times L$ lattices at the 
percolation density. Fig.~\ref{fig}a shows results for the disorder-averaged 
staggered structure factor (the equal-time spin correlation function at 
wave-vector ${\bf q}=(\pi,\pi)$, defined as in \cite{kato}) versus inverse 
temperature for three different system sizes ($J$ is the Heisenberg coupling).
The same configurations of magnetic sites were used at all temperatures
and the statistical errors are therefore correlated and completely dominated 
by fluctuations between different site configurations (several hundred random 
site configurations were used for each $L$). In \cite{kato} lattices with 
$L \le 48$ were studied at temperatures down to $J/1000$. Fig.~\ref{fig}a
reveals that for $L=48$ the structure factor has not yet converged to its 
asymptotic $T=0$ value at this temperature. A careful statistical analysis, 
taking into account the covariance between data at different temperatures, 
gives that $S(\pi,\pi)$ at $J/T=1024$ is $(95.0 \pm 0.5)\%$ of the almost
saturated value at $J/T = 8192$. It is thus likely that the finite-size 
scaling in \cite{kato} is to some extent affected by temperature.

Temperature effects alone cannot explain the anomalous behavior of the 
critical exponents found for small $S$ in \cite{kato}. In order to better 
understand the scaling behavior of $S(\pi,\pi)$ it is useful to also study 
the cluster structure factor $S_{\rm cl}(\pi,\pi)$ defined on the largest 
connected cluster of magnetic sites. The corresponding cluster magnetization 
is given by $m^2_{\rm cl} = \langle 3S_{\rm cl} (\pi,\pi)/N_{\rm cl} \rangle$,
where $N_{\rm cl}$ is the size of the cluster and $\langle~\rangle$ denotes 
the average over random site configurations. In the thermodynamic limit the 
clusters have fractal dimension $D=91/48$ and in \cite{kato} it was argued 
that the spin correlations on these clusters decay algebraically with an 
$S$-dependent exponent. Fig.~\ref{fig}b shows $m^2_{\rm cl}$ versus 
$1/L^{D/2}$ along with a quadratic fit. The cluster magnetization clearly 
scales to a finite value as $L \to \infty$ and therefore, contrary to the 
claim of \cite{kato}, the fractal clusters for $S=1/2$ are long-range ordered.
The scaling of the full $S(\pi,\pi)$ must then be classical and, as was also
concluded in \cite{kato}, the critical density $p_{\rm c} \equiv p^*$. 

The reason why the classical exponents are not seen in the full staggered 
structure factor considered in \cite{kato} is that the cluster magnetization 
decays considerably with increasing lattice size for the sizes used, as can 
be seen in Fig.~\ref{fig}b. This partially compensates for the growth of 
the fractal clusters with  $L$, leading to a slower growth of $S(\pi,\pi)$
with $L$ than in a classical system for which $m^2_{\rm cl}$ is constant. 
The asymptotic behavior can be observed only when $L$ is sufficiently large 
for the relative size corrections to $m^2_{\rm cl}$ to be small. This quantum 
mechanical effect makes it extremely difficult to directly observe the true 
critical behavior for small values of $S$.

Support by NSF grant DMR-97-12765 is acknowledged. The simulations were 
carried out on the NCSA Origin2000 system and the Condor system at the
University of Wisconsin -- Madison. 
\null

\end{document}